# Lowest Degree Decomposition of Complex Networks

Yong Yu[1], Ming Jing[1], Na Zhao[1,2], Tao Zhou[3]

[1] Key Laboratory in Software Engineering of Yunnan Province, School of Software, Yunnan University, Kunming 650091, P. R. China
[2] Electric Power Research Institute of Yunnan Power Grid, Kunming 650217, P.R. China
[3] CompleX Lab, University of Electronic Science and Technology of China, Chengdu 611471, P. R. China

**The heterogeneous structure [1,2] implies that a very few nodes may play the critical role in maintaining structural and functional properties of a large-scale network. Identifying these vital nodes is one of the most important tasks in network science [3,4], which allow us to better conduct successful social advertisements [5], immunize a network against epidemics [6], discover drug target candidates and essential proteins [7], and prevent cascading breakdowns in power grids [8], financial markets [9] and ecological systems [10]. Inspired by the nested nature of real networks [11], we propose a decomposition method where at each step the nodes with the lowest degree are pruned. We have strictly proved that this so-called lowest degree decomposition (LDD) is a subdivision of the famous k-core decomposition [12,13]. Extensive numerical analyses on epidemic spreading, synchronization and nonlinear mutualistic dynamics show that the LDD can more accurately find out the most influential spreaders, the most efficient controllers and the most vulnerable species than k-core decomposition and other well-known indices [14-17]. The present method only makes use of local topological information, and thus has high potential to become a powerful tool for network analysis.**

Given a network $G(V,E)$ where $V$ and $E$ are the sets of nodes and links, the problem of vital nodes identification can be formulated in three related yet different ways [4]: (i) to find out the minimal set $V'\subset V$ that satisfies some certain requirements (e.g., the feedback vertex set problem [18]); (ii) to determine the set $V'\subset V$ with a preset size $|V'|=n$, which has the maximal dynamical impact (e.g., the influence maximization problem [19]); (iii) to provide a ranking of nodes' functional significances based on their structural features, which is usually on the basis of a proper centrality measure. Accordingly, many centralities have been developed to characterize influences of individual nodes, ranging from simple measures like degree [1], closeness [20] and betweenness [16], to elaborately designed ones like PageRank [21], LeaderRank [22] and collective influence [23].

A particularly interesting group of methods is network decomposition, with an underlying hypothesis that nodes are organized in different levels and the ones at the nucleus are most influential, which is to some extent supported by recent empirical evidences about the nested organization [11] and core-periphery structure [24]. Aiming at identifying influential nodes, the most successful decomposition method till far is the k-core decomposition [14]. Considering a connected simple network $G$ where multiple links and self-loops are not allowed, the k-core decomposition process starts by removing all nodes with degree $k=1$. This causes new nodes with degree $k \leq 1$ to appear. These are also removed and the process stops when all remaining nodes are of degree $k>1$. The removed nodes and their associated links form the 1-shell, and the nodes in the 1-shell are assigned a k-shell value $k_s$=1. This pruning process is repeated to extract the 2-shell, that is, in each step the nodes with degree $k \leq 2$ are removed. Nodes in the 2-shell are assigned a k-shell value $k_s$=2. The process is continued until all higher-layer shells have been identified and all nodes have been removed. Recent empirical and theoretical studies [10,14] both suggest that the k-shell index is a good measure of a node's influence: a higher $k_s$ indicates a larger influence.

A severe drawback of the k-core decomposition is that $k_s$ is not sufficiently distinguishable as each shell may contain numerous nodes. Some modified methods are recently proposed, mainly via replacing degree in the decomposition process by other centralities or combining k-shell index with other centralities (see such variants of the k-core decomposition in Ref. [4]). These modifications bring some certain improvement in accuracy, together with complicated details that clouds our understanding about network organization. Here we propose a even simpler decomposition method named as lowest degree decomposition (LDD for short). Firstly, the nodes with the lowest degree are removed, which form the 1-shell under LDD and are assigned a value $L_s$=1. Then, the remaining nodes with the lowest degree are removed, which form the 2-shell with $L_s$=2. This pruning process stops when all nodes have been removed. A notable difference from k-core decomposition is that LDD peels off every shell at once, without any iterations.

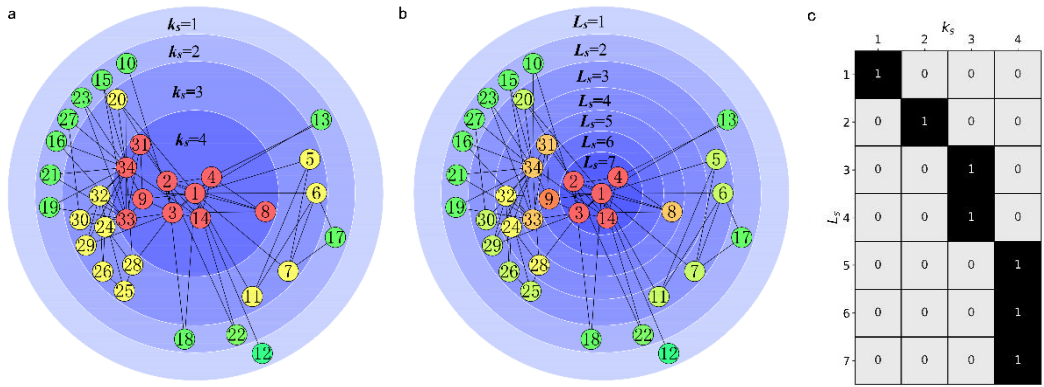

**Fig. 1 | Illustration of k-core decomposition and lowest degree decomposition**. **a**, A schematic representation of the Zachary karate club network under the k-core decomposition, where the maximum value of k-shell index is $k_s^{\max}=4$. **b**, A schematic representation of the Zachary karate club network under the lowest degree decomposition, where the maximum value of LDD index is $L_s^{\max}=7$. **c**, The associated matrix $M$ of $k_s$ and $L_s$, where the entities marked by dark black are of value 1 and the entities colored by light grey are of value 0.

Taking the well-studied Zachary karate club network [25] as an example, figure 1 illustrates the results from k-core decomposition (Fig. 1a) and LDD (Fig. 1b). It seems that LDD is more distinguishable as $L_s$ has 7 different values ($L_s^{\max}=7$) while $k_s$ has only 4 ($k_s^{\max}=4$). To reveal the relationship between LDD and k-core decomposition, we propose a simple binary matrix $\Lambda \subset \{0,1\}^{L_s^{\max} \times k_s^{\max}}$, named as associated matrix, where the element $\Lambda_{uv} = 1$ if there exists at least one node $i$ satisfying that $L_s(i) = u$ and $k_s(i) = v$, and $\Lambda_{uv} = 0$ otherwise. Figure 1c shows the associated matrix for the Zachary karate club network, from which one can observe that for each $k_s$ value, there can be one or more associated $L_s$ values, whilst each $L_s$ value is only associated with one $k_s$ value. This is not a coincidence. Indeed, we have strictly proved that LDD is a subdivision of k-core decomposition, that is, nodes in one k-shell can be assigned by different $L_s$ values, while nodes with the same $L_s$ value must belong to one k-shell. In addition, given any two nodes $i$ and $j$, if $k_s(i) > k_s(j)$, then $L_s(i) > L_s(j)$, and if $L_s(i) > L_s(j)$, then $k_s(i) \geq k_s(j)$. The proof is presented in the Supplementary Section I.

We further improve the resolution of LDD by utilizing the neighborhood information. Denote $N_i$ the number of nodes whose $L_s$ values are smaller than $L_s(i)$, $N$ the number of nodes in the target network, and $S_i = N_i/N$ the ratio of nodes $i$ can beat subject to LDD, we propose a so-called LDD+ index for any node $i$ as

$$L_s^+(i) = S_i + \alpha \sum_{j \in \Gamma_i} S_j,$$

which takes into account the importance of $i$'s neighbors. The free parameter $\alpha$ is set to balance the contributions from $i$ itself and its neighbors. We can also define the k-shell+ index in a similar way. Notice that, though later we will show that LDD+ and k-shell+ indices perform better than LDD and k-shell indices, we do not think the former are better than the latter since we have to tune one parameter in LDD+ and k-shell+ indices, while LDD and k-shell indices are parameter-free.

To see whether LDD can be used to characterize individual nodes' influences, we use 9 real networks from disparate fields for experimental analyses. They are all simple networks, where directionality and weight of any link are ignored and self-loops are not allowed. In brief, there are one word network (AdjNoun), two communication networks (Email-Enron and Email-URV), two biological networks (PPI and Enzyme), two social networks (Dublin and Hamsterster), and two power grids (Bcs and Ops). Detailed descriptions, corresponding references and topological statistics of these networks are presented in Supplementary Section II. We compare the proposed methods, LDD and LDD+, with six benchmark indices including k-shell [14], k-shell+, degree, betweenness [16], H-index [15], and mixed degree decomposition (MDD) [17]. The precise definitions of betweenness, H-index and MDD are shown in Methods. In the later experimental analyses, if an index contains a tunable parameter, its value will be turned to the one corresponding to the best performance.

We first test whether LDD can well quantify a node's influence in spreading dynamics by applying two standard spreading model, the susceptible-infected-recovered (SIR) model and the susceptible-infected-susceptible (SIS) model [26] (see Methods for the model descriptions). For SIR model, the influence of a node $i$, say $R_i$, is defined as the number of eventually recovered nodes averaged over 1000 independent runs, each of which starts with node $i$ being the sole infected seed. For SIS model, the influence of a node $i$, say $P_i$, is defined as the probability that node $i$ is infected at time $t$ when $t \to \infty$. In simulation, $P_i$ is obtained by averaging over 1000 independent runs, and in each run averaging over 100 time steps after the system reaches a dynamic-equilibrium state, where as many infected nodes become susceptible as susceptible nodes become infected. Given $L_s$ of all nodes as $L_s(i)$, $i = 1, 2, \cdots, N$, and the node influences for SIR and SIS by simulation as $R_i$ and $P_i$, respectively, we apply the Kendall's Tau ($\tau$) [27], namely $\tau(L_s, R)$ and $\tau(L_s, P)$, to quantify to what extent $L_s$ resembles spreading influences of individual nodes. The values of $\tau$ lies in the range $-1 \leq \tau \leq 1$, and the larger value means a stronger correlation (see Methods for the definition of $\tau$). The performance of other indices under consideration is also measured by $\tau$.

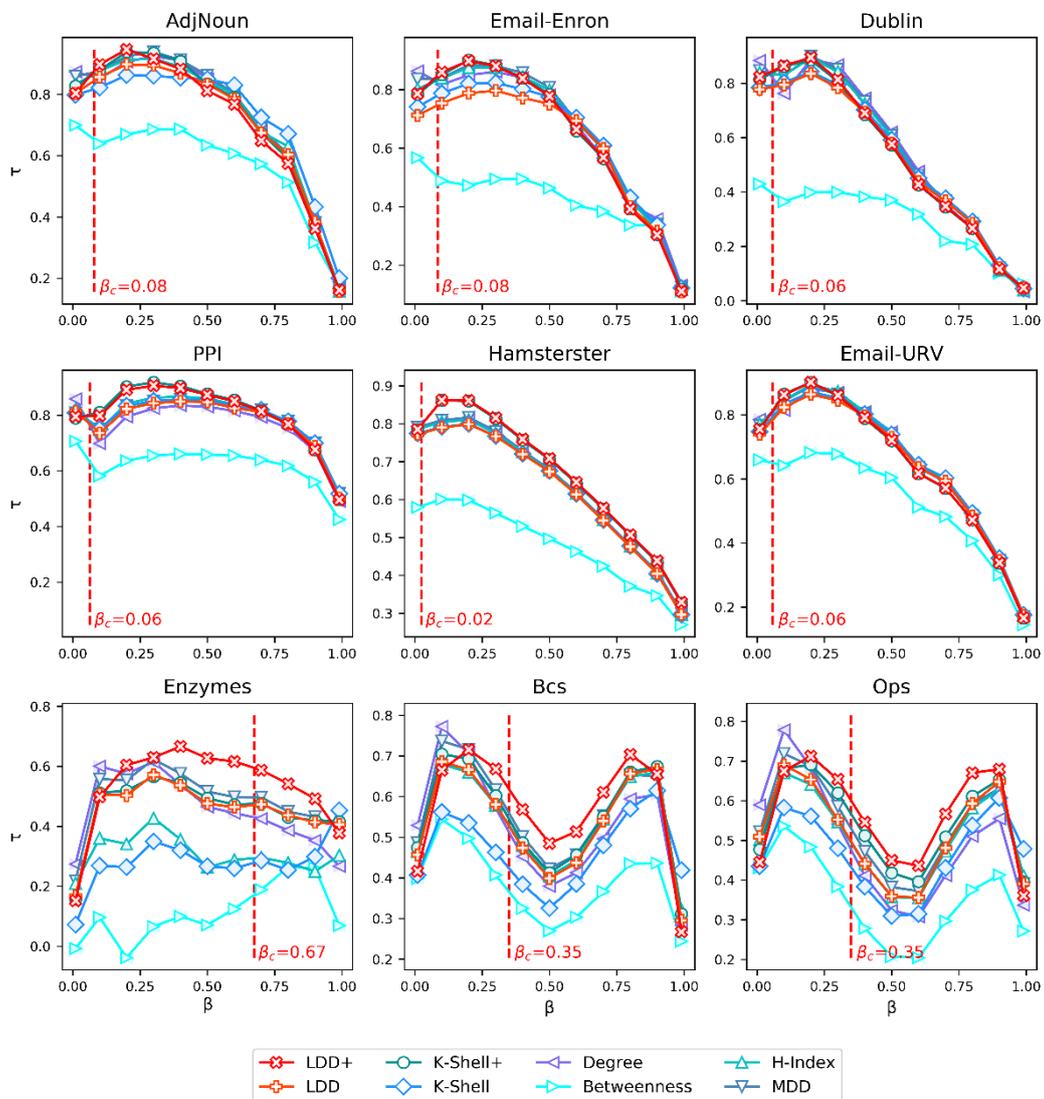

**Fig. 2 | Performance of the eight indices in evaluating nodes' spreading influences according to the SIR model.** Each plot represents a real network, with the vertical dash line marking the epidemic threshold. The results of LDD and LDD+ are emphasized by red color.

Figure 2 compares the eight indices on the nine real networks for the full spectrum of $\beta$ under the SIR model. One can observe that the $\tau-\beta$ curves of LDD and LDD+ are usually on top, suggesting their superiority in identifying influential spreaders. In the SIR model, when $\beta$ is very small, the disease cannot spread out and the infected node only has a small chance to infect its immediate neighbors, so that the problem to estimate a node's spreading influence becomes trivial and the best index is just the number of neighbors, say degree. In contrast, when $\beta$ is very high, the disease will infect a large percentage of the population, irrespective of where it originated, and thus the individual influence is meaningless. Accordingly, we mainly focus on the range around the epidemic threshold [28]. Table 1 reports the values of $\tau(L_s, R)$ for all considered indices at the threshold of each network. One can observe that LDD+ and k-shell+ perform much better than other indices, and subject to the average value of $\tau$, LDD+ is slightly better than k-shell+.

**Table 1 | The performance of the 8 indices on 9 networks for the SIR model at the epidemic threshold. The best-performed results are emphasized in bold.**

| Networks | LDD+ | K-Shell+ | LDD | K-Shell | Degree | Betweenness | H-Index | MDD |
|---|---|---|---|---|---|---|---|---|
| AdjNoun | **0.885** | 0.872 | 0.843 | 0.805 | 0.859 | 0.662 | 0.860 | 0.871 |
| Email-Enron | **0.858** | 0.854 | 0.755 | 0.789 | 0.820 | 0.480 | 0.844 | 0.847 |
| Dublin | 0.862 | **0.865** | 0.805 | 0.812 | 0.770 | 0.376 | 0.834 | 0.837 |
| PPI | 0.778 | **0.789** | 0.728 | 0.736 | 0.701 | 0.590 | 0.737 | 0.736 |
| Hamsterster | 0.755 | **0.756** | 0.722 | 0.724 | 0.733 | 0.566 | 0.731 | 0.735 |
| Email-URV | 0.760 | **0.761** | 0.739 | 0.747 | 0.737 | 0.608 | 0.756 | 0.755 |
| Enzyme | **0.603** | 0.477 | 0.475 | 0.278 | 0.420 | 0.193 | 0.286 | 0.494 |
| Bcs | **0.631** | 0.553 | 0.534 | 0.420 | 0.515 | 0.368 | 0.537 | 0.565 |
| Ops | **0.613** | 0.579 | 0.507 | 0.441 | 0.487 | 0.340 | 0.502 | 0.542 |
| Average | **0.749** | 0.723 | 0.679 | 0.639 | 0.672 | 0.465 | 0.676 | 0.709 |

For SIS model, in each run, 20% of randomly selected nodes are initially set to be infected and others are susceptible. Since local hubs can self-sustain the spreading, the predicted epidemic thresholds based on mean-field approximation [29] usually deviate from the true thresholds of real networks. Therefore, for the nine real networks under consideration, we use simulations to determine the thresholds $\beta_c$. In practice, given a network, we start with $\beta=0.01$ and increase the value of $\beta$ by a step length 0.01. The numerically estimated $\beta_c$ is the first value in the $\beta$ sequence such that the system reaches the dynamic-equilibrium state in each of 1000 independent runs. Generally speaking, the estimated threshold is slightly above the true value, but as the resolution is 0.01, the deviation is very small. Table 2 reports the values of $\tau(L_s, P)$ for all considered indices at the threshold of each network. Similar to the result of SIR model, LDD+ and k-shell+ perform much better than other indices, and subject to the average value of $\tau$, LDD+ is slightly better than k-shell+.

Table 2 | **The performance of the 8 indices on 9 networks for the SIS model at the epidemic threshold. The best-performed results are emphasized in bold.**

| Networks | LDD+ | K-Shell+ | LDD | K-Shell | Degree | Betweenness | H-Index | MDD |
|---|---|---|---|---|---|---|---|---|
| AdjNoun | **0.949** | 0.933 | 0.890 | 0.847 | 0.898 | 0.667 | 0.907 | 0.919 |
| Email-Enron | 0.910 | **0.913** | 0.790 | 0.824 | 0.851 | 0.470 | 0.883 | 0.893 |
| Dublin | **0.903** | 0.902 | 0.820 | 0.828 | 0.789 | 0.365 | 0.858 | 0.858 |
| PPI | 0.804 | **0.816** | 0.736 | 0.749 | 0.697 | 0.580 | 0.749 | 0.749 |
| Hamsterster | 0.874 | **0.875** | 0.789 | 0.791 | 0.806 | 0.612 | 0.806 | 0.814 |
| Email-URV | 0.906 | **0.908** | 0.845 | 0.854 | 0.836 | 0.655 | 0.870 | 0.871 |
| Enzyme | **0.646** | 0.510 | 0.499 | 0.290 | 0.470 | 0.158 | 0.295 | 0.540 |
| Bcs | **0.537** | 0.452 | 0.429 | 0.326 | 0.405 | 0.270 | 0.432 | 0.461 |
| Ops | **0.483** | 0.450 | 0.368 | 0.314 | 0.341 | 0.214 | 0.370 | 0.408 |
| Average | **0.779** | 0.751 | 0.685 | 0.647 | 0.677 | 0.443 | 0.686 | 0.724 |

Next, we test whether LDD can dig out influential nodes in a synchronizing process subject to their controlling efficiency under pinning control [30,31], which is an effective method to drive the system from any initial state to a targeted synchronized state (see Methods for the description of pinning control and how to measure individual nodes' influences). As shown in Table 3, LDD+, k-shell+ and MDD are competitive to each other, and perform better than the other five indices.

Table 3 | **The performance of the 8 indices on 9 networks for the pinning control. The best-performed results are emphasized in bold.**

| Networks | LDD+ | K-Shell+ | LDD | K-Shell | Degree | Betweenness | H-Index | MDD |
|---|---|---|---|---|---|---|---|---|
| AdjNoun | 0.935 | 0.951 | 0.894 | 0.856 | 0.957 | 0.711 | 0.916 | **0.964** |
| Email-Enron | 0.829 | 0.833 | 0.724 | 0.753 | **0.881** | 0.581 | 0.822 | **0.881** |
| Dublin | 0.869 | 0.860 | 0.805 | 0.812 | 0.882 | 0.441 | 0.869 | **0.889** |
| PPI | 0.942 | **0.947** | 0.861 | 0.864 | 0.861 | 0.681 | 0.197 | 0.883 |
| Hamsterster | **0.920** | 0.918 | 0.845 | 0.846 | 0.896 | 0.560 | 0.394 | 0.901 |
| Email-URV | 0.945 | 0.939 | 0.886 | 0.887 | 0.950 | 0.723 | 0.929 | **0.958** |
| Enzyme | **0.545** | 0.483 | 0.493 | 0.360 | 0.367 | 0.366 | 0.299 | 0.501 |
| Bcs | **0.581** | 0.560 | 0.535 | 0.482 | 0.501 | 0.448 | 0.537 | 0.540 |
| Ops | **0.558** | 0.523 | 0.498 | 0.458 | 0.434 | 0.389 | 0.497 | 0.499 |
| Average | **0.791** | 0.780 | 0.727 | 0.702 | 0.748 | 0.545 | 0.607 | 0.779 |

Lastly, we consider a dynamical process in a mutualistic ecosystem consisting of $N$ interacting species, which can be represented by a network of $N$ nodes, each of which corresponds to a species. Two species $i$ and $j$ share a link if $i$ interacts with $j$, with the interacting strength being denoted by $\gamma_{ij} > 0$. The dynamical model [10,32] defines how the density of each species $i$, $x_i(t)$, evolves according to its current density and the interactions with species (see Methods for details). For simplicity, we set $\gamma_{ij} = \gamma$. By gradually turning down $\gamma$ from a large initial value, we can find the threshold $\gamma_c$ by simulation, at or below which all species will die out. It is expected that a species in a more central place of the network is easier to survive [10], so we fix $\gamma = \gamma_c$ and record the order of extinctions of all species, then we calculate the Kendall's Tau between the ranking by $L_s$ (from large to small) and the inverse ranking of extinctions. We compare performance of the 8 indices based on 27 real ecological mutualistic networks (details about these networks are presented in Supplementary Section III). As shown in Table 4, LDD+ performs best among all considered indices, and LDD performs better than k-shell index. Therefore, LDD can be used as a powerful tool to identify the most vulnerable species in a mutualistic ecosystem.

**Table 4 | | The performance of the 8 indices on 9 networks to predict the extinction order for ecological mutualistic dynamics. The best-performed results are emphasized in bold.**

| Networks | LDD+ | K-Shell+ | LDD | K-Shell | Degree | Betweenness | H-Index | MDD |
|---|---|---|---|---|---|---|---|---|
| 1 | **0.821** | 0.802 | 0.756 | 0.756 | 0.692 | 0.598 | 0.750 | 0.699 |
| 2 | **0.849** | 0.764 | 0.725 | 0.678 | 0.664 | 0.644 | 0.688 | 0.678 |
| 3 | **0.687** | 0.564 | 0.585 | 0.554 | 0.540 | 0.464 | 0.573 | 0.544 |
| 4 | **0.843** | 0.807 | 0.803 | 0.781 | 0.771 | 0.739 | 0.796 | 0.782 |
| 5 | **0.829** | 0.795 | 0.705 | 0.615 | 0.750 | 0.684 | 0.669 | 0.752 |
| 6 | **0.771** | 0.707 | 0.678 | 0.666 | 0.608 | 0.580 | 0.683 | 0.615 |
| 7 | **0.804** | 0.794 | 0.719 | 0.734 | 0.663 | 0.610 | 0.718 | 0.674 |
| 8 | **0.796** | 0.778 | 0.763 | 0.763 | 0.730 | 0.658 | 0.744 | 0.731 |
| 9 | **0.810** | 0.810 | 0.761 | 0.762 | 0.758 | 0.723 | 0.785 | 0.766 |
| 10 | **0.788** | 0.744 | 0.618 | 0.633 | 0.580 | 0.547 | 0.635 | 0.585 |
| 11 | **0.920** | 0.897 | 0.902 | 0.865 | 0.853 | 0.749 | 0.912 | 0.863 |
| 12 | **0.814** | 0.769 | 0.691 | 0.703 | 0.642 | 0.624 | 0.701 | 0.649 |
| 13 | **0.632** | 0.579 | 0.515 | 0.549 | 0.458 | 0.419 | 0.546 | 0.475 |
| 14 | **0.886** | 0.870 | 0.721 | 0.727 | 0.706 | 0.688 | 0.724 | 0.709 |
| 15 | **0.859** | 0.850 | 0.775 | 0.782 | 0.734 | 0.656 | 0.784 | 0.742 |
| 16 | **0.802** | 0.793 | 0.639 | 0.674 | 0.606 | 0.573 | 0.668 | 0.616 |
| 17 | **0.730** | 0.708 | 0.597 | 0.621 | 0.554 | 0.515 | 0.617 | 0.564 |
| 18 | **0.823** | 0.733 | 0.582 | 0.591 | 0.566 | 0.560 | 0.588 | 0.568 |
| 19 | **0.817** | 0.784 | 0.690 | 0.708 | 0.656 | 0.611 | 0.703 | 0.661 |
| 20 | **0.940** | 0.908 | 0.759 | 0.775 | 0.716 | 0.675 | 0.781 | 0.735 |
| 21 | 0.874 | 0.858 | 0.857 | 0.849 | 0.844 | 0.839 | **0.888** | 0.858 |
| 22 | 0.895 | 0.893 | 0.870 | 0.751 | 0.892 | 0.821 | 0.809 | **0.927** |
| 23 | 0.825 | 0.785 | 0.779 | 0.647 | 0.760 | 0.689 | 0.647 | **0.843** |
| 24 | **0.876** | 0.820 | 0.804 | 0.499 | 0.770 | 0.724 | 0.765 | 0.815 |
| 25 | **0.840** | 0.807 | 0.783 | 0.772 | 0.730 | 0.601 | 0.791 | 0.741 |
| 26 | 0.864 | 0.876 | **0.885** | 0.836 | 0.840 | 0.757 | 0.883 | 0.843 |
| 27 | **0.902** | 0.899 | 0.819 | 0.690 | 0.857 | 0.717 | 0.863 | 0.857 |
| **Average** | **0.826** | 0.792 | 0.733 | 0.703 | 0.702 | 0.647 | 0.730 | 0.715 |

In summary, according to extensive experiments n representative network-based dynamical processes, LDD+ show best ability to identify influential nodes, and LDD performs better than k-shell index. Therefore, we conclude that the lowest degree decomposition is an effective method to unfold hidden information of networks. We have implemented robust analyses (see results in Supplementary Section IV), suggesting that the advantages of the lowest degree decomposition still hold for other normal settings of dynamical parameters. Scientists have already proposed some variants of k-core decomposition by introducing additional parameters and more complicated operations, or by directly combining k-core decomposition with other centrality indices. Overall speaking, those methods sacrifice elegance for better performance. In contrast, as a novel decomposition method, LDD is even simpler than k-core decomposition and the strong mathematical tie between LDD and k-core decomposition is very clear. Analogous to k-core decomposition, LDD can also be extended to deal with directed networks and weighted networks. In addition to the identification of influential nodes, LDD can also be used as a powerful toll for network visualization and as a criterion to validate the network evolution models.

## Methods

**Kendall's Tau**. We consider any two indices associated with all $N$ nodes, $X = (x_1, x_2, \cdots, x_N)$ and $Y = (y_1, y_2, \cdots, y_N)$, as well as the $N$ two-tuples $(x_1, y_1), (x_2, y_2), \cdots, (x_N, y_N)$. Any pair $(x_i, y_i)$ and $(x_j, y_j)$ are concordant if the ranks for both elements agree, namely if both $x_i > x_j$ and $y_i > y_j$ or if both $x_i < x_j$ and $y_i < y_j$. They are discordant if $x_i > x_j$ and $y_i < y_j$ or if $x_i < x_j$ and $y_i > y_j$. If $x_i = x_j$ or $y_i = y_j$, the pair is neither concordant or discordant. Comparing all $N(N-1)/2$ pairs of two-tuples, the Kendall's Tau is defined as

$$\tau(X,Y) = \frac{2(n_+ - n_-)}{N(N-1)},$$

where $n_+$ and $n_-$ are the number of concordant and discordant pairs, respectively. If $X$ and $Y$ are independent, $\tau$ should be close to zero, and thus the extent to which $\tau$ exceeds zero indicates the strength of correlation.

**Benchmark Indices**. Betweenness Centrality of a node $i$ is defined as

$$BC(i) = \sum_{s \neq i, s \neq t, i \neq t} \frac{g_{st}(i)}{g_{st}},$$

where $g_{st}(i)$ is the number of shortest paths between nodes $s$ and $t$ which pass through node $i$, and $g_{st}$ is the number of all shortest paths between nodes $s$ and $t$. H-index of a node $i$, denoted as $H(i)$, is defined as the maximal integer satisfying that there are at least $H(i)$ neighbors of node $i$ with the degrees no less than $H(i)$. The MDD algorithm defines a so-called mixed degree $k_i^{(m)}$ for each node $i$. Initially, $k_i^{(m)}$ is set to be $i$'s degree $k_i$. The procedure of MDD includes three steps as follows. Step (i): The nodes with minimum mixed degree (denoted by $M_{\min}$) are removed and added to the $M_{min}$-shell. Step (ii): Each remaining node $i$'s mixed degree is updated as $k_i^{(m)} \leftarrow k_i^{(r)} + \mu k_i^{(e)}$, where $k_i^{(r)}$ is the number of $i$'s original neighbors in the remaining network, $k_i^{(e)}$ is the number of $i$'s original neighbors having been removed, and $\mu$ is a tunable parameter in the range [0,1]. If some nodes' updated mixed degrees are no more than $M_{min}$, they will be removed and added to the $M_{min}$-shell, and Step (ii) will be repeated until all remaining nodes' mixed degrees are larger than $M_{min}$. All removed nodes constitute the $M_{min}$-shell and are assigned a MDD value $M_{min}$. Step (iii): Repeat Steps (i) and (ii) until the remaining network is empty.

**SIR model**. Nodes in a networked SIR model can be in one of three possible states: susceptible, infected and recovered. The SIR process begins with one or more infected seeds and all other nodes are initially susceptible. At each time step, each infected node contacts its neighbors and each susceptible node has an infectivity probability $\beta$ to be infected by one infected neighbor. Then, each previously infected node enters the recovered state with a probability $\gamma$. We set $\gamma = 1$ for simplicity. According to the heterogeneous mean-field theory [28], the epidemic threshold of SIR model is approximate to

$$\beta_c \approx \frac{\langle k \rangle}{\langle k^2 \rangle - \langle k \rangle},$$

where $\langle k \rangle$ and $\langle k^2 \rangle$ denote the mean degree and mean square of degree.

**SIS model**. The SIS model describes the spreading processes of a large number of infections that do not confer full immunity to infected individuals after infections. In a networked SIS model, a node can be susceptible or infected. Initially, a number of nodes are set to be infected seeds and others are susceptible. At each time step, each infected node contacts its neighbors and each susceptible node has an infectivity probability $\beta$ to be infected by one infected neighbor. Then, each infected node returns to the susceptible state with probability $\lambda$. In the experimental analyses, we fix $\lambda = 0.1$.

**Pinning Control**. Considering a general case where a simple connected network $G$ is consisted of $N$ coupled nodes, with interacting dynamics as

$$\dot{x}_i = f(x_i) + \sigma \sum_{j=1}^{N} l_{ij} \Gamma(x_j) + U_i(x_1, \ldots, x_N),$$

where the vector $x_i \in \mathbf{R}^n$ is the state of node $i$, the function $f(\cdot)$ describes the self-dynamics of a node, the positive constant $\sigma$ denotes the coupling strength, $U_i$ is the controller applied at node $i$, and the inner coupling matrix $\Gamma: \mathbf{R}^n \to \mathbf{R}^n$ is positive semidefinite. The Laplacian matrix of $G$ is $L = D - A$, where $D$ and $A$ are respectively the degree matrix and adjacency matrix of $G$, and $l_{ij}$ is the element in $L$. The controlling efficiency of an arbitrary node $i$ can be measured by the smallest nonzero eigenvalue $\mu_1(L_i)$, where $L_i$ is the principle submatrix obtained by deleting the $i$th row and column of $L$ [30,31].

**Ecological Mutualistic Dynamics.** The networked mutualistic dynamics can be described a set of nonlinear differential equations with $i=1,2,\ldots,N$ [10,32]:

$$\frac{\partial x_i(t)}{\partial t} = -dx_i - sx_i^2 + \sum_{j=1}^{N} a_{ij}\gamma_{ij} \frac{x_i x_j}{\alpha + \sum_{k=1}^{N} a_{ik}x_k},$$

where $d>0$ is the death rate, $s>0$ is the self-limitation parameter encoding the competition of resources that limits a species' growth, $a_{ij}$ is the element of the adjacency matrix $A$, $\gamma_{ij}$ is the interacting strength between species $i$ and $j$, $\alpha$ is the half-saturation coefficient. In the experimental results reported in Table 4, we set the initial density for every species $i$ as $x_i(0)=1$, and fix $d=0.2$, $s=1$ and $\alpha=1$. If a species' density has decreases to 0, it is considered to be extinct, so the extinction order of all species can be obtained by numerical simulation.

# References

1. Barabási, A.-L. & Albert, R. Emergence of Scaling in Random Networks. *Science* **286**, 509–512 (1999).
2. Caldarelli, G. *Scale-Free Networks: Complex Webs in Nature and Technology* (Oxford University Press, 2007).
3. Pei, S. & Makse, H. A. Spreading dynamics in complex networks. *J. Stat. Mech. Theory Exp.* P12002 (2013).
4. Lü, L., Chen, D.-B., Ren, X.-L., Zhang, Q.-M., Zhang, Y.-C. & Zhou, T. Vital nodes identification in complex networks. *Phys. Rep.* **650**, 1–63 (2016).
5. Leskovec, J., Adamic, L. A. & Huberman, B. A. The dynamics of viral marketing. *ACM Trans. Web* **1**, 5 (2007).
6. Chen, Y., Paul, G., Havlin, S., Liljeros, F. & Stanley, H. E. Finding a Better Immunization Strategy. *Phys. Rev. Lett.* **101**, 058701 (2008).
7. Csermely, P., Korcsmáros, T., Kiss, H. J. M., London, G. & Nussinov, R. Structure and dynamics of molecular networks: A novel paradigm of drug discovery. *Pharmacol. Ther.* **138**, 333–408 (2013).
8. Albert, R., Albert, I. & Nakarado, G. L. Structural vulnerability of the North American power grid. *Phys. Rev. E* **69**, 025103 (2004).
9. Haldane, A. G. & May, R. M. Systemic risk in banking ecosystems. *Nature* **469**, 351–355 (2011).
10. Morone, F., Del Ferraro, G. & Makse, H. A. The k-core as a predictor of structural collapse in mutualistic ecosystems. *Nat. Phys.* **15**, 95–102 (2019).

11. Mariani, M. S., Ren, Z.-M., Bascompte, J. & Tessone, C. J. Nestedness in complex networks: Observation, emergence, and implications. *Phys. Rep.* **813**, 1–90 (2019).

12. Seidman, S. B. Network structure and minimum degree. *Soc. Networks* **5**, 269–287 (1983).

13. Dorogovtsev, S. N., Goltsev, A. V. & Mendes, J. F. F. k-Core Organization of Complex Networks. *Phys. Rev. Lett.* **96**, 040601 (2006).

14. Kitsak, M., Gallos, L. K., Havlin, S., Liljeros, F., Muchnik, L., Stanley, H. E. & Makse, H. A. Identification of influential spreaders in complex networks. *Nat. Phys.* **6**, 888–893 (2010).

15. Lü, L., Zhou, T., Zhang, Q.-M. & Stanley, H. E. The H-index of a network node and its relation to degree and coreness. *Nat. Commun.* **7**, 10168 (2016).

16. Freeman, L. C. A Set of Measures of Centrality Based on Betweenness. *Sociometry* **40**, 35-41 (1977).

17. Zeng, A. & Zhang, C.-J. Ranking spreaders by decomposing complex networks. *Phys. Lett. A* **377**, 1031-1035 (2013).

18. Zhou, H.-J., Spin glass approach to the feedback vertex set problem, *Eur. Phys. J. B* **86**, 455 (2013).

19. Kempe, D., Kleinberg, J. & Tardos, E. Maximizing the spread of influence through a social network. In *Proc. 9th ACM SIGKDD Int. Conf. on Knowledge Discovery and Data Mining*, 137-143 (ACM, 2003).

20. Bavelas, A. Communication patterns in task-oriented groups. *J. Acoust. Soc. Am.* **22**, 725-730 (1950).

21. Brin, S. & Page, L. The anatomy of a large-scale hypertextual web search engine. *Comput. Networks ISDN Syst.* **30**, 107-117 (1998).

22. Lü, L., Zhang, Y.-C., Yeung, C. H. & Zhou, T. Leaders in Social Networks, the Delicious Case. *PLoS ONE* **6**, e21202 (2011).

23. Morone, F. & Makse, H. A. Influence maximization in complex networks through optimal percolation. *Nature* **524**, 65-68 (2015).

24. Csermely, P., London, A., Wu, L.-Y. & Uzzi, B. Structure and dynamics of core/periphery networks. *J. Complex Networks* **1**, 93-123 (2013).

25. Zachary, W. W. An Information Flow Model for Conflict and Fission in Small Groups. *J. Anthropol. Res.* **33**, 452–473 (1977).

26. Pastor-Satorras, R., Castellano, C., Van Mieghem, P. & Vespignani, A. Epidemic processes in complex networks. *Rev. Mod. Phys.* **87**, 925–979 (2015).

27. Kendall, M. A new measure of rank correlation. *Biometrika* **30**, 81-89 (1938).

28. Castellano, C. & Pastor-Satorras, R. Thresholds for epidemic spreading in networks. *Phys. Rev. Lett.* **105**, 218701 (2010).

29. Pastor-Satorras, R. & Vespignani, A. Epidemic spreading in scale-free networks. *Phys. Rev. Lett.* **86**, 3200-3203 (2001).

30. Li, X., Wang, X. & Chen, G. Pinning a complex dynamical network to its equilibrium. *IEEE Trans. Circuits Syst. I* **51**, 2074-2087 (2004).

31. Pirani, M. & Sundaram, S. On the Smallest Eigenvalue of Grounded Laplacian Matrices. *IEEE Trans. Automat. Contr.* **61**, 509–514 (2016).

32. May, R. M. Mutualistic interactions among species. *Nature* **296**, 803-804 (1982).

## Acknowledgements

This work is partially supported by the National Natural Science Foundation of China under Grant No. 61433014, the Innovation Team Project of Yunnan Province under No. 2017HC012, and the National Key Research and Development Program under No. 2018YFB2100100.

## Author contributions